# The Chimaerical Quest for the Optical Plasmonic Superlens

George Christou and Christos Mias[1§]

[1]School of Engineering, Warwick University, Coventry, CV4 7AL, UK

SYNOPSIS

This commentary aims to expose the fallacy of claiming that a plasmonic silver film superlens is capable to image real subwavelength objects. This lens was proposed by the Berkeley's group who, in their misleading experiment, inappropriately regarded subwavelength apertures as the objects to be imaged whereas the main function of these apertures was to transform free space laser light into an evanescent field necessary for exciting the surface plasmon resonance phenomenon in silver. In addition, the apertures also determined the constrained effective area on the silver film where the phenomenon could occur. We provide a fresh insightful physical explanation of how this phenomenon is excited and what it entails. We emphasize the phenomenon's important effect of subwavelength conversion (reduction) of the generated surface plasmons and their associated bound enhanced evanescent fields.

[§] Corresponding author: c.mias@warwick.ac.uk, tel: 00 44 2476522343, fax: 00 44 2476418922.
Submitted: 4 April 2014.

The aspiration to use visible light radiation to image a subwavelength object led to the concept of the negative refraction optical superlens. Eventually, the plasmonic silver film superlens was proposed based on the surface plasmon resonance phenomenon but associated with negative refraction. In [1] we attempted to dissociate the natural surface plasmon resonance phenomenon from the concept of negative refraction. Nevertheless, Pendry mentions in [2] that the Xiang Zhang group at the University of California, Berkeley, demonstrated a lens that could resolve details as small as one-sixth the wavelength of visible light. The above statement is fallacious. What this lens really does is to reduce the size of the details of an "object". This issue will be critically investigated and clarified by a fresh insightful consideration of the effects of the surface plasmon resonance phenomenon.

The lens, Pendry was referring to, was the well known plasmonic superlens, the silver film superlens [3]. Briefly, the Berkeley's group performed an experiment in which the surface plasmon resonance phenomenon was excited by an evanescent field of 365-nm-wavelength. This evanescent field resulted from the passage of 365-nm-wavelength laser light through shaped apertures of 40-nm width in a 50-nm thick opaque chromium mask. Using this method to generate the evanescent field was cunning because the Berkeley's group could also present these apertures as the subwavelength objects that are being imaged. In fact the results they obtained were acclaimed to be a proof for the capability of the silver film superlens to image subwavelength objects and, most amazingly, as a proof for the soundness of the superlens concept. We shall challenge their interpretation of their results and we shall expose the fallacy of claiming subwavelength imaging with the so-called silver film superlens.

We first consider briefly how we can excite, photonically, the fundamentally important surface plasmon resonance phenomenon. For this phenomenon to occur there must be an energy and momentum matching between the incident photons and the free conduction electrons of the silver film. But we know that at the natural resonant frequency of silver the free space photons have a small momentum due to the very small relativistic mass whereas the electrons have a larger momentum as a result of their rest mass. Fortunately, this mismatch can be bridged by transforming propagating waves into evanescent waves which have a greater momentum. With this elucidating information we revisit the Berkeley's group experiment. We emphasise that the function of the 40-nm wide apertures is to transform some of the 365-nm-wavelength ultraviolet laser light into an evanescent wave, of the same wavelength, necessary and indispensable for this lens to achieve a resonance at the natural resonant frequency of silver. Achieving this resonance generates 60-nm-wavelength surface plasmon waves which set-up, bound to them, enhanced vertical evanescent fields on the image side of the silver film. It is to be noted that the wavelength value of the generated surface plasmons was not verified experimentally by us. However, as this value is only used for illustration purposes it is acceptable by us.

Now, how can we explain, physically, why for this lens the incident photons of the evanescent waves of 365-nm-wavelength generate surface plasmon waves of 60-nm-wavelength? It can be simply explained if we recall that the electrons have a much shorter wavelength than the photons [1] and consequently the wavelength of the electrons' collective oscillations – the surface plasmon waves – is affected diminishingly. Also, how short the wavelength of the surface plasmon waves becomes, additionally depends on the physical and geometrical parameters of the materials being used. Whereas in reference [1] this large wavelength difference between the photons and the electrons helped us to understand the increase in the field intensity enhancement, in this paper, this significant difference between their wavelengths helps us to understand the wavelength reduction of the recorded enhanced evanescent field. The essence of what we strive to stress is that the surface plasmon resonance phenomenon converts an evanescent field of 365-nm-wavelength into an enhanced evanescent field of 60-nm-wavelength. This is conceptually illustrated in Figure 1. It is this conversion (reduction) in wavelength that we are witnessing rather than the resolution of an "object's subwavelength details" claimed to be as small as one-sixth the wavelength of visible light - we should not forget that the 365-nm wavelength evanescent field is the "object" that is being imaged. This conversion in wavelengths, due to the natural silver film resonance, can be exploited in lithography to fabricate nanoscale structures much smaller than the wavelength of the illuminating laser light. Also, this reduction in wavelength of the surface plasmons is certainly the basis of useful applications in the field of plasmonics. However, we are doubtful if this reduction in wavelength can be of any use for the imaging of real subwavelength objects and it is misleading to present this wavelength conversion as subwavelength imaging.

Next we consider another role of the subwavelength apertures apart from their function of transforming propagating waves into evanescent waves. The area of the apertures imposes a constraint upon the effective area on the silver film surface where the surface plasmon resonance phenomenon is allowed to occur. Consequently, there is an analogous area constraint for the photoresist where the concentrated energy of the converted in wavelength evanescent field is recorded. We therefore stress that the area constraint in combination with the wavelength reduction of the enhanced evanescent field, determine the subwavelength effect seen by the Berkeley's group and erroneously considered by them as proof for subwavelength imaging with the silver film.

To settle the issue of the subwavelength imaging capability of the so-called silver film superlens it is appropriate for the Berkeley's group, or for any other superlens proponents, to accept the following challenge in the imaging of real subwavelength objects. To image either a subwavelength solid object of 40-nm diameter having a coin-like surface or viruses which have a subwavelength size of about 100-nm. Otherwise we are justified to claim that a working silver film superlens, the poor man's superlens, has been a chimaera.

We conclude this critique by emphasising that the offered physical explanation of the wavelength conversion is based on the fundamentally important photon-electron energy interaction that takes place at the natural resonant frequency of silver. We also remind the reader that the Berkeley's group experiment is an application of the natural surface plasmon resonance phenomenon in silver, phenomenon which is not at all associated with the concept of negative refraction. Additionally, by applying Occam's Razor (William of Occam 1290-1350) we declare that there are no grounds for assuming anything unphysical in the various applications of the surface plasmon resonance phenomenon, an entirely physical phenomenon.

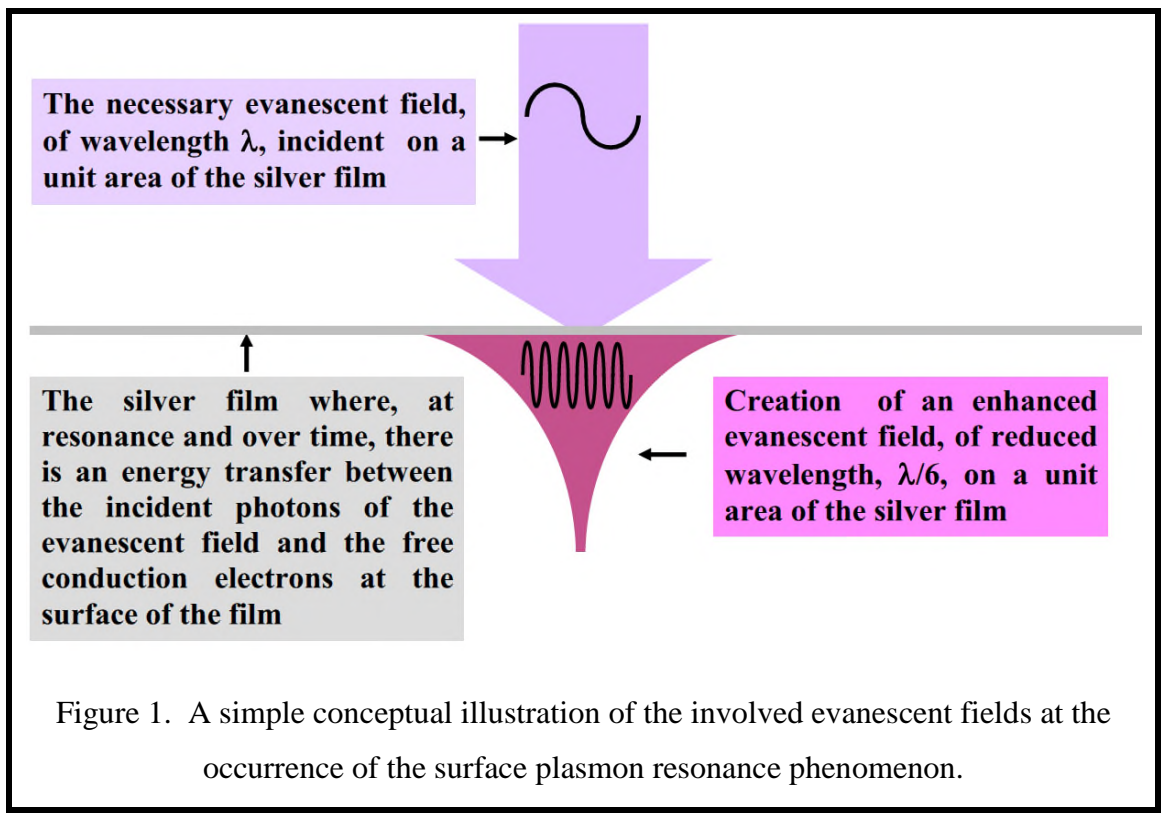

Figure 1. A simple conceptual illustration of the involved evanescent fields at the occurrence of the surface plasmon resonance phenomenon.